# Ticket to Ride: Impact of free public transport on women's workforce participation in India

Udayan Rathore
(SJMSoM, IIT Bombay; Wadhwani Foundation. E-mail: rathore.udayan@gmail.com )

Ashish Singh
(SJMSoM, IIT Bombay. E-mail: singh.ashish@iitb.ac.in )

*This version: 10$^{th}$ February, 2026*

## Abstract

*We leverage a quasi-natural experiment from India on introduction of free bus schemes for women across five states to study its impact on women's workforce participation. We use two rounds of the representative Time Use Survey (2019 and 2024) and a triple difference estimation strategy, complemented by an event study framework to identify the causal relationship of interest. Findings reveal that the bus scheme was successful in improving women's paid work participation and duration of employment. We confirm that these results are not merely a continuation of prior trends. The scheme's effects are concentrated among early adopters like Punjab and Tamil Nadu, two states with historically different levels of women's workforce participation. We also find disproportionately higher effects for women residing in more patriarchal districts with higher mobility restrictions. We argue that the scheme works through easing of non-financial binding constraints, which lowers barriers to women's mobility and workforce participation.*

Word Count (excluding references and annex): 7,000
Number of Exhibits (excluding annex): 5
Declarations of Interest: None

Corresponding Author and Address: Udayan Rathore, A-128, Sector-21, Jal Vayu Vihar, NOIDA, UP, India (201301), rathore.udayan@gmail.com

Acknowledgements: I am thankful to Nandini Jayakumar and Shreya Bhattacharya for feedback on the initial version of the manuscript.

1. **Introduction**

Improving women's labour force participation (WLFP) is intrinsically and instrumentally important for a variety of reasons (Heath, et al., 2024). In addition to enhanced household income, WLFP is positively associated with higher economic growth (Bell, et al., 2019; Hsieh, Hurst, Jones, & Klenow, 2019), improved human capital formation (Atkin, 2009; Heath & Mobarak, 2015; Qian, 2008) and women empowerment (Majlesi, 2016). The latter are in-turn related to improvements in multiple other measures of human development (Balasubramanian, Ibanez, Khan, & Sahoo, 2024). Higher WLFP is thus considered crucial for achieving United Nation's (UN) Sustainable Development Goals (SDGs), especially SDG 5 (Gender equality) and SDG 8 (Decent work and economic growth) (Verick, 2018). These factors notwithstanding, global WLFP continues to lag work participation rates for men with particularly large disparities in South Asia and India (World Bank, 2025).

Transportation is a crucial factor that influences labour force participation (Gutiérrez-i-Puigarnau & van Ommeren, 2010). In contrast to men however, women's travel patterns are more non-linear and tedious due to disproportionately higher burden of chores and care work, which involves multiple interconnected trips or "trip-chaining" (Perez, 2019). Thus, women's mobility patterns across the world are closely intertwined with availability of, and access to public transport, which is a crucial supply side factor that influences women's workforce participation (Alam, Cropper, Dappe, & Suri, 2021; Perez, 2019). Evidence from developed countries shows that free public women transit schemes increase ridership, they have little effect on workforce participation (Ofosu-Kwabe, Lim, & Malalgoda, 2024). However, WLFP is not a function of travel costs alone but is also constrained by socio-cultural norms, safety concerns and lack of decision-making by women (Borker, 2022; Heath, et al., 2024).[1] This question assumes particular importance for low and middle income countries (LMIC), as it sits at the intersection of low disposable income, increasing safety concerns for women, stagnating WLFP and limited government resources.

In this context, we study the effects of a quasi-natural experiment from India, where free bus schemes for women were introduced across five states between 2019 and 2023. We ask three primary questions of interest. One, can free public transport for women increase their workforce participation and extend the time they spend in paid work? Two, are these effects uniform, or do they vary across states and by duration of exposure to the scheme? Third, does the impact of the scheme vary with the intensity of local patriarchal norms and mobility restrictions?

In this paper, we harmonise state-level variation in rollout of free bus schemes for women with the two rounds of the Time Use Surveys (TUS, 2019 and 2024) and the fourth round of the Demographic and Health Survey (DHS, 2016) to study the relationships of interest. Our identification strategy is based on a triple-difference estimator, which compares change in workforce participation on the extensive margin (any paid work) and the intensive margin (minutes of paid work) by gender, over states that implemented a free bus scheme for women and compare them against states that did not, both before and after the policy roll-out. We also use the event study framework to account for prior trends.

---

[1] There is a technical distinction between labour force and workforce. The former also includes those who are currently not working but seeking work (unemployed). Given that we are using time use information on paid work undertaken, when we mention labour force, we are referring to workforce.

We find that the free bus schemes have a positive effect on WLFP, both on the extensive and the intensive margins. We confirm that these improvements can be attributed to the scheme rollout and is not a continuation of prior trends. Our findings also provide evidence in support this mechanism by revealing that the scheme supported a significant increase in probability of women travelling for work. We also find that the bus scheme facilitates reallocation of time towards socialisation and leisure, without increasing the burden of unpaid domestic labour. We argue that this may support creation of social capital that might be instrumental in easing socio-cultural barriers faced by women. Our results show that the effects of the scheme vary significantly by states. These effects are more pronouced among early adopters of the scheme like Punjab and Tamil Nadu (2021), two states which are on opposite ends of the WLFP spectrum in India. Finally, we see that the effects of the bus scheme are diproportionately better for women residing in districts that experience higher socio-cultural barriers to women's mobility and participation in paid work. s

Our research contributes to the emerging literature on the role of public-transport in facilitating women's mobility and improving their workforce participation. Evidence shows that improved transport infrastructure, such as Rapid Bus Transit (BRT), elevated light rails systems, road networks, frequent services and last mile connectivity can support women's employment and earnings (Alam, Cropper, Dappe, & Suri, 2021; Borker, 2024; Lei, Desai, & Vanneman, 2019; Martinez, et al., 2020). Various studies show primacy of "safe space" during travel (Borker, 2024; Kondylis, et al., 2020). Although technology enabled ride-hailing services tackle some of these concerns, they are exclusionary as their benefits accrue mostly to the economically well-off sections of society (Christensen & Osman, 2023). Similarly, while provisions for exclusive transport for women are effective (Aguilar, Gutiérrez, & Villagrán, 2021; Field & Vyborny, 2022), they introduce a cost burden, where resources must be duplicated to maintain reliable frequency for both women-only and other travelers. In comparison, free transit policies for women with sufficiently high ridership allows for economies of scale and promotes collective surveillance and mutual safety. The effectiveness of such schemes however need to be rigorously evaluated to inform optimal policy designs.

Our paper makes several important contibutions to the literature. One, using state representative datasets, we find that the free bus scheme for women has a dual effect where more women participate in any paid work (extensive margin) for durations (intensive margin). Thus, the scheme not only broadens women's participation but also deepens this engagement. Two, we find that the effects of the free bus scheme are disproprtionately stronger for women in districts that experience higher patriarhcy and mobility restrictions. We argue that beyond lowering the travel cost, the effectiveness of the scheme is driven by easing non-financial binding constraints on WLFP. We suspect that increase in women's ridership and shared occupancy, which takes time to realise, strengthens '*safety in numbers*' and crowds-out the threat of harassment through mutual surveillance. The precise mechanisms of how the bus scheme interacts with socio-cultural norms is beyond the scope of this paper and constitutes questions for future research. Three, in the Indian context, our research complements two pieces of ongoing research on this topic in India. In their ongoing work, Dasgupta & Datta (2023) study the impact of Delhi's free bus scheme for women, launched on October 29, 2019, on WLFP. The study uses the TUS-2019 round and finds weak evidence of the scheme improving women's workforce participation. As the survey covers only the calendar year of 2019, the analysis is constrained by both a small sample size of respondents who had access to the scheme and the duration of exposure. In their working paper, Chen, et al., (2024) primarily use the Consumer Pyramids Household Survey (CPHS) to study the impact of the bus schemes in two states (Punjab and Tamil Nadu), and find that the scheme does not have a uniform effect on WLFP, with labour supply increasing only for skilled women. These findings however need to be interpreted with caution as evidence shows that the CPHS suffers from "main street" bias, which underrepresents poor and marginalised households, even after reweighting (Dreze & Somanchi, 2021; Roy & Van Der Weide, 2025; Somanchi, 2021). Notably, CPHS's estimates on WLFP diverge systematically from the nationally representative Periodic Labour Force Survey (PLFS), conducted by the National Sample Survey Office (NSSO), which is India's leading authority on representative sample surveys (Abraham & Shrivastava, 2022). In contrast, our analysis is based on the three representative rounds of survey, two from the TUS and one from the DHS and covers all five states where bus schemes was launched between 2019 and 2024. This allows us to comprehensively examine the effects of the policy on WLFP.

The remainder of the paper is structured as follows. Section 2 provides a brief background of the free bus schemes for women implemented across different states. Section 3 presents the data and methodology. Section 4 summarises our findings and Section 5 concludes with a discussion.

## 2. Background

On October 29th, 2019, the Government of National Capital Territory of Delhi implemented India's first Free-Fare Public Transport (FFPT) scheme for women, which fully subsidised bus rides on all the state transport corporations' buses (Jamba, Devaraj, & Kanuri, 2025). The scheme aims to promote women's mobility thereby fostering women's empowerment and safety, through increased ridership and other safety features (Aam Aadmi Party, 2019). Between its inception and January 2023, government figures show that about 1000 million free tickets were issued (Dasgupta & Datta, 2023). However, report from the Comptroller and Auditor General (CAG), India's foremost auditing authority reveals that the fleet size of state operated buses declined by more than 13 percent between 2015 and 2022, and almost half of all low-floor buses were in disrepair, suffering frequent breakdowns (Rathore & Barnagarwala, 2025). This ties with the evidence of overcrowding, unreliable scheduling, and rise in safety concerns for women, which may have offset the intended goal of promoting mobility and facilitating higher WLFP (Jamba, Devaraj, & Kanuri, 2025).

The scheme garnered widespread interest with multiple state governments introducing similar schemes for women. In April 2021 and May 2021, Government of Punjab (Free Bus Travel Scheme) and Government of Tamil Nadu ('*Magalir Vidiyal Payanam*' or Journey of Dawn for Women) respectively, rolled out free bus schemes for women. Preliminary evidence from Tamil Nadu shows that the scheme was associated with higher household savings, greater frequency of travel, particularly among women from marginalised groups (State Planning Commission, 2022). Following suit, the government of Karnataka and Telangana rolled out the *'Shakti'* and *'Mahalakshmi'* schemes in June and December of 2023, respectively. The latter is a more comprehensive scheme which also offers other guarantees like subsidised clean cooking fuel, unconditional cash transfer in addition to free rides for women on state transport buses. Most recently, the Government of Andhra Pradesh implemented the '*Stree Shakti*' scheme in August of 2025 with similar objectives, a period beyond the scope of the current study.

Except Andhra Pradesh, where the scheme was introduced by the Telugu Desam Party (TDP) who are in alliance with the Bharatiya Janata Party (BJP) and part of the ruling National Democratic Alliance (NDA) coalition, all other state governments are part of the Indian National Developmental Inclusive Alliance (INDIA) coalition the which is currently in the opposition. The political debates on the effectiveness of the scheme has been viewed from the lens of fiscal conservatism at one end, and welfare based entitlements on the other. The former focusses on rising state government debts, which may crowd out private investment in transport infrastructure that has become increasingly important for State Transport Undertakings (STUs) under the Public-Private Partnership (PPP) model. The latter emphasises the rights-based approach and the social value created through expanding women's access to and participation in public spaces, education and paid work. Although reports show an increase in women's ridership numbers, little is known on whether the bus scheme directly impacted women's workforce participation. Our research bridges this gap by rigorously evaluating this relationship of interest. We now discuss the data and methodology used in this paper.

## 3. Data and Methodology
### 3.1 Data

We use two consistent rounds of the TUS conducted by NSSO that are representative at the state level. Both surveys have an identical sampling strategy which follows a stratified, two stage random sampling design. The First Stage Units (FSUs) are selected from villages or Urban Frame Survey (UFS) and at the second stage, the Ultimate Stage Units (USUs) are sample households from rural or urban sector. Sample stratification is based on district, village and town sizes (Ministry of Statistics and Programme Implementation (MOSPI), 2025; Rathore, 2026). The TUS have a distinct advantage over traditional labour market surveys, which use the major time criteria to identify the activity on which the individual spent most of their time during a reference period, usually past year (Usual Principal Activity Status, PS) or week (Current Weekly Status, CWS). This approach has been shown to underestimate workforce participation due to misclassification, particularly for women who often undertake multiple tasks simultaneously (Hirway & Jose, 2011).

These two rounds of TUS were conducted between January and December, first in 2019 and then in 2024. They use a 24-hour recall, starting at 4:00 AM on the day before the interview, continuing up to 4:00 AM on the date of the survey. The survey covers information on 165 potential activities as per the structure aligned with the International Classification of Activities for Time-Use Statistics (ICATUS 2016). The activities are recorded in up to 48 thirty-minute windows for household members that are six years or older. The survey allows for up to three simultaneous activities, classifies them into major or minor categories and verifies if they were undertaken inside or outside the premises. The 2019 and 2024 rounds cover 10,004 and 10,024 FSUs and 138,799 and 139,487 households, respectively. We restrict our analysis to working age members between the ages of 18 and 60 years. In addition to the time-use information, the surveys also collect detailed socio-economic information on household members. We also use the fourth round of the representative DHS from 2015-2016, which follows a stratified, multi-stage, sampling design. This is used to create a district level India Patriarchy Index (IPI) based on the methodology proposed by Singh, et al., (2021) and capture mobility measures for women.

*3.2 Variables of interest*

For participation in paid work, we use two primary outcome variables. The TUS surveys capture time spent on employment related activities, which can be major or minor. We consider only major activities to define and construct the variables for any paid work participation (extensive margin) and total duration of paid work undertaken (intensive margin). The TUS also records time spent on employment related travel or commute.[2] We use this information to create similar intermediate outcomes on the extensive and intensive margins, respectively. The variable however does not record the mode of travel. Nonetheless, if the bus scheme was successful in promoting WLFP, it should translate to improvement in these overall metrics. We are also interested in how the time saved, if any, is reallocated amongst different activities like socialisation, leisure and unpaid household labour.

There are three main independent variables of interest, which are as follows: (a) Gender, which takes a value of 1 for women and 0 otherwise; (b) Spatial exposure to bus scheme for women, which takes a value of 1 for respondents from the five states where such a scheme was rolled out, 0 otherwise[3]; and (c) Pre-Post variable that takes a value of 0 before policy rollout (2019) and 1 afterwards (2024). Apart from New-Delhi, four other states implemented the scheme in 2021 and 2023, between the two TUS rounds. For New-Delhi, the pre-post variable takes a value of 1 from October 29, 2019, which is the date of scheme roll-out.[4] We use a range of individual and household level control variables. The former includes age, its square, education level, and marital status, while latter covers household level demography, settlement type (rural or urban), socio-religious affiliation and economic status. We restrict the analysis to normal days where the individual pursued their routine activities.[5] The survey also records respondents' state and district of residence, alongside survey dates. To maintain comparability of administrative jurisdictions, we include all states and the National Capital Territory (NCT) of New-Delhi but exclude six Unition Territories (UTs) from the analysis. Unlike the former group, which have elected governments, UTs are directly administered by the Union Government.[6] We also compile state wise, public sector bus fleet data for the period 2017-2018 to 2024-2025 using multiple government sources (Association of State Road Transport Undertakings (ASRTU), 2024; Ministry of Road Transport & Highways (MoRTH), 2020, 2023).[7] We also use Census 2011 data to compute population serviced per bus from above to account for public bus availability across states. For heterogeneity analysis, we use the representative DHS-4 data to construct a district representative patriarchy index (IPI), as used in Singh, et al., (2021) and Rathore (2026). The IPI is a composite index that is aggregated across 12 indicators of patriarchy across five domains. The domains include male domination over women, generational dominance, patrilocality, son preference and socio-economic domination. Similarly, we create a district level indicator of barriers to women's mobility based on share of women who are not allowed to visit the market (MR).

---

[2] The ICATUS codes differentiate between employment related travel (181) and commute (182).

*3.3 Empirical strategy*

To study the relationship of interest, we use two rounds of representation TUS data from India and exploit the introduction of free bus scheme for women across five states as a quasi-natural experiment to study its impact on women's workforce participation. The effects of the scheme are identified through a Triple Difference (DDD) design. This facilitates a comparison between men and women, in states that implemented the free bus scheme for women against the states that did not, over a period when the schemes were in effect versus when they were not. This is detailed in equation (1) below.

---

[3] The variable also takes a value of 0 for Andhra Pradesh where scheme was introduced only in 2025, which is beyond the period covered in the data (2019-2024).
[4] In robustness checks, we confirm that findings are not sensitive to this coding scheme.
[5] Alternatively, days of weekly offs, leaves, social functions or illness are classified as '*other-days*'.
[6] UTs include Goa, Puducherry, Andaman and Nicobar Islands, Lakshadweep, Chandigarh and Dadra and Nagar Haveli & Daman and Diu.
[7] The conceptualisation of this dataset, the identification of relevant sources, and analytical decisions for calculations were made by the authors. The dataset was compiled and cleaned with the assistance of Claude (Sonnet 4.6 model). Results are robust to exclusion of this variable. A summary of how this data was constructed and assumptions made is available in Supplementary material, Section-S1.

$$Y_{ihst} = \beta_0 + \beta_1 Women_{ihst} + \beta_2 Post_t + \beta_3 BusStates_s + \beta_4(Women_{ihst} * Post_t) + \beta_5(Women_{ihst} * BusStates_s) + \beta_6(Post_t * BusStates_s) + \beta_7(Women_{ihst} * Post_t * BusStates_s) + X'_i\gamma + X'_h\delta + State_s + Year_t + (State_s * Year_t) + \varepsilon_{ihst} \quad (1)$$

$Y_{ihst}$ refers to the outcomes (Section 3.2) of individual *'i'*, in household *'h'*, in state *'s'* in year *'t'*. The outcomes are regressed on gender ($Women_{ihrst}$), treatment status of states ($BusStates_s$), for periods before and after scheme rollout ($Post_t$), alongside their two and three-way interactions. We also account for a range of individual and household level control variables that are represented by vectors $X'_i$ and $X'_h$, respectively. To account for time-invariant unobservables for states, broad country-level time trends, we include state and year fixed effects. Additionally, we also include State*Year fixed effects, which accounts for any state specific shocks that affected men and women in a similar way.[8] The standard errors are clustered two-way by state and quarters. This accounts for correlations within state across time, and those within common quarters, irrespective of states.[9] $\beta_7$ above is the parameter of interest which informs the Average Treatment Effects on the Treated (ATT) for women, assuming common trends assumption holds.

Additionally, to check for differential pre-trends, we also undertake event study estimation, which utilizes exogeneity of survey timing across the two rounds of the TUS. The NSSO uses a multistage, stratified random sampling design, where the randomly selected FSUs are uniformly allocated to each of the sub-round to ensure even spread of sampled FSUs over the entire study period. Thus, the households have no control over the timing of survey which can be considered as exogeneous. In equation (1) above, we replace $Post_t$ dummy variable with $Event\_Quarter_t = k, for\ k = 1,2,...8$.[10] This represents a set of quarter indicators for the two years of analysis. We consider the fourth quarter by calendar (Q4, 2019) as the base category, as this was the last period before the rollout of bus schemes in all the treatment states except New-Delhi. As the scheme in New-Delhi was introduced on 29th October 2019, well into Q4 of 2019, we use two empirical approaches to ensure that our results are not confounded by this timing. First, we exclude the sample from New-Delhi entirely to maintain a consistent pre-treatment baseline. Second, we retain this sample but allocate everyone who were interviewed after the scheme rollout to the subsequent treatment period, while those who were interviewed before this date remain in the baseline period. We hypothesise that the treatment effects remained constant before the policy rollout. If the scheme was effective in improving WLFP, we expect to see a significant and sustained divergence in WLFP in the subsequent quarters.[11]

---

[8] We note that the effects of $BusStates_s$ and ($Post_t * BusStates_s$) are absorbed by State and State*Year FEs. Also note that State*Year FEs absorbs the state level variation in population serviced per public sector bus, which we use as a control.
[9] For estimation, we use the reghdfe routine on Stata 17.
[10] Note that values 1 to 4 correspond to each quarter in 2019 where 5 to 8 for those in 2024.

Since the five states rolled out the schemes at different points of time, one in 2019; two in 2021 and two in 2023; the heterogeneous treatment effects estimated for each state confound the state level characteristics with duration of exposure. This cannot be disentangled and requires data from additional time periods. Nonetheless, to explore this combined heterogeneity across states, we replace the binary $BusStates_s$ variable in equation (1) above with a set of treatment state-specific dummies. To study how the scheme interacts with norms and its effects on WLFP, we run two separate restricted models for working-aged women where we focus on triple interaction of treatment status ($BusStates_s$), before and after the scheme ($Post_t$) with two variants of standardised district level indicators. These are (a) Patriarchy ($zIPI_d$), and (b) Restrictions in visiting the market ($zMR_d$). In the estimation equation (2) below, we denote these by $zNorms_d$. The specification in comparable to equation 1 above, with the subscript 'd' denoting district level indicators. $\gamma_7$ is the coefficient of interest and informs the differentiated effects of the scheme on workforce participation for women residing in districts characterised by higher patriarchy and mobility restrictions. Next, we check discuss the findings and see if the bus scheme had any impact on WLFP.

$$Y_{ihsdt} = \gamma_0 + \gamma_1 zNorms_d + \gamma_2 Post_t + \gamma_3 BusStates_s + \gamma_4(zNorms_d * Post_t) + \gamma_5(zNorms_d * BusStates_s) + \gamma_6(Post_t * BusStates_s) + \gamma_7(zNorms_d * Post_t * BusStates_s) + X'_i\lambda + X'_h\mu + State_s + Year_t + (State_s * Year_t) + \varepsilon'_{ihst} \quad (2)$$

## 4. Results

*4.1 Summary Statistics*

The all-India share of working-age women in paid work was 23 percent in 2019. The five states that implemented free bus scheme for women are distributed on either side of this mean value. While New-Delhi and Punjab were marginally above the first quartile (Q1) and reported WLFP rates of about 17 percent and 16 percent, Karnataka stood at 27 percent (Q3). In comparison, for Tamil Nadu and Telangana these rates were about 39 percent (Q1) and 47 percent (Q1), respectively. By 2024, the all-India WLFP estimate increased to more than 27 percent. Although New-Delhi and Punjab continued to be in the lower half of the distribution, women's workforce participation improved, particularly in Punjab where this increased by about 9 pp. to 26 percent. On the other end of the distribution, these rates also improved for Karnataka (33 percent) and Tamil Nadu (42 percent), while remaining at similar levels for Telangana (Annex, TableA1).

These improvements were accompanied by a general increase in work force participation, which was higher for men. As a result, the all-India gender gap has widened from 56 percent in 2019 to

---
[11] For robustness, we also use a higher frequency specification with months instead of quarters.

59 percent in 2024. In 2019, this gap was particularly high for New-Delhi and Punjab at about 67 percent in 2019, which worsened for the former (74 percent), but improved for the latter (62%). For Tamil Nadu and Telangana, this remained comparable at about 48 percent and 41 percent, respectively (Annex, TableA1). We now present the results from multivariate regression analysis, where we estimate the impact of the free bus scheme for women on their workforce participation.

*4.2 Regression Results*

*a. Main Results: Impact on WLFP*

The main set of results from our triple-difference model are presented in Table-1 below. We are primarily interested in the coefficient of the triple interaction term, which identifies the ATT estimates of the free bus scheme. We estimate these effects both on the extensive (any paid work) and the intensive margins (minutes of paid work) of WLFP. We first present the former where we add different set of covariates in stages to verify the stability of the relationship of interest. Model-1 includes the three variables of interest, their two-way and three-way interactions alongside state and year fixed effects. In models 2 and 3, we additionally control for individual and household level covariates. Model-4 represents the full-specification and also includes state-year fixed effects. Model-5 presents the full-specification estimates for WLFP on the intensive margin. To verify the plausibility of our main findings, we evaluate the bus scheme's impact on commuting behaviour. Specifically, we estimate the impact of the scheme on the probability of any paid-work related commute (Model-6) and the average minutes spent on such commute (Model-7) (Table-1).

Results in column 4 indicate that the free public bus scheme increased WLFP by about 3 percentage points (pp.) (p-value <0.01) in comparison to the reference group.[12] On the intensive margin (column 5), we find that the average daily work duration increased by about 16 minutes a day (p-value <0.10). Although these coefficients may appear modest, they represent a dual impact of broader workforce participation among women alongside an increase in daily work duration. To put these effect sizes in context, in 2019 about 24 percent of working-age women were engaged in paid work with an average duration of 86 minutes a day. The results represent an average increase of about 13 percent in workforce participation on the extensive margin and an average expansion of 20 percent in the work duration on the intensive margin. A favourable transport scheme that improves WLFP is expected to increase the share of women engaged in work related travel or commute. Our findings support this mechanism and reveal a 5 pp. increase (p-value <0.01) in the likelihood of women reporting any work-related travel (Table-1).

**Table-1:** Average Treatment Effects on Treated (ATT) effects on workforce participation

| | *Any time on paid work/Extensive Margin (=1)* | *Mins. on paid work /Intensive* | *Any time on paid work-* | *Mins. on paid work- related* |
|---|---|---|---|---|

---

[12] The reference group is men in non-scheme states before the roll-out.

|  | Model-1 | Model-2 | Model-3 | Model-4 | Margin Model-5 | related travel (=1) Model-6 | travel Model-7 |
|---|---|---|---|---|---|---|---|
|  | (1) b/se | (2) b/se | (3) b/se | (4) b/se | (5) b/se | (6) b/se | (7) b/se |
| Women (Base=Men) | -0.562*** | -0.580*** | -0.578*** | -0.578*** | -313.64*** | -0.452*** | -40.064*** |
|  | (0.021) | (0.021) | (0.021) | (0.021) | (8.952) | (0.019) | (1.957) |
| Post (Base=Pre) | 0.029* | 0.036** | 0.032** | 0.002 | -22.849*** | 0.038*** | -0.853 |
|  | (0.014) | (0.013) | (0.012) | (0.004) | (3.412) | (0.010) | (1.603) |
| Women # Post | -0.035*** | -0.032*** | -0.032*** | -0.033*** | -22.836** | -0.075** | -3.636 |
|  | (0.006) | (0.008) | (0.009) | (0.009) | (6.982) | (0.023) | (2.193) |
| Women # Bus Scheme States | 0.028 | 0.023 | 0.023 | 0.023 | -13.932 | -0.041** | -5.121** |
|  | (0.036) | (0.036) | (0.036) | (0.036) | (17.795) | (0.017) | (1.762) |
| Women x Post x Bus Scheme States | 0.026** | 0.027** | 0.027** | 0.028*** | 15.614* | 0.048*** | 3.345 |
|  | (0.009) | (0.009) | (0.009) | (0.008) | (7.625) | (0.013) | (2.000) |
| Constant | 0.813*** | -0.009 | 0.002 | 0.017 | 12.114 | 0.033 | 2.666 |
|  | (0.011) | (0.060) | (0.058) | (0.059) | (29.947) | (0.058) | (5.371) |
| Individual Controls | ✗ | ✓ | ✓ | ✓ | ✓ | ✓ | ✓ |
| Household Controls | ✗ | ✗ | ✓ | ✓ | ✓ | ✓ | ✓ |
| State FE | ✓ | ✓ | ✓ | ✓ | ✓ | ✓ | ✓ |
| Year FE | ✓ | ✓ | ✓ | ✓ | ✓ | ✓ | ✓ |
| (State#Year) FE | ✗ | ✗ | ✗ | ✓ | ✓ | ✓ | ✓ |
| Constant | 0.813*** | -0.009 | 0.002 | 0.017 | 12.114 | 0.033 | 2.666 |
|  | (0.011) | (0.060) | (0.058) | (0.059) | (29.947) | (0.058) | (5.371) |
| Observations | 571,099 | 571,099 | 571,096 | 571,096 | 571,096 | 571,099 | 571,099 |
| $R^2$ | 0.353 | 0.399 | 0.400 | 0.402 | 0.447 | 0.303 | 0.224 |

Notes: Author's calculations, *** p-value < 0.01, ** p-value < 0.05, * p-value <0.10, standard errors are clustered two-way, by state and quarters. Model-1 controls the main three variables of interest and their two and three-way interactions, alongside State and Year fixed effects. Model-2 additionally controls individual-level covariates. Model-3 adds household-level controls. Model-4 is the full model and adds State # Year fixed effects. Coefficients of "Bus scheme States" and "Bus scheme States * Post" are subsumed under State and State*Year fixed effects, respectively.

### b. Event Study: Checking for prior trends

The primary threat to our identification comes from the possibility of these effects representing a continuation or prior trends rather than the effects of the bus scheme. We use an event study framework to formally test the validity of the common trends assumption for the results on any workforce participation. Following the empirical strategy discussed in Section 3.3, we estimate a series of triple interaction coefficients for different calendar quarters. This allows us to trace the trajectory of WLFP in treated states in comparison to the reference category, before and after the bus policy implementation.[13]

In Figure-1, the coefficients for three quarters preceding the policy rollout are statistically indistinguishable from zero. This result confirms that the overall results are not driven by any divergent pre-trends before the introduction of the policy. The treatment effects are positive and significant for the three quarters observed post policy period (2024), while the estimate for the fourth quarter is positive but insignificant. The findings are consistent for both the specifications of either excluding New-Delhi to maintain a clean baseline (Panel A) or including it and reallocating respondents to quarters based on the scheme rollout date (Panel B). The results are robust to a higher frequency specification, which uses months instead of quarters in the specification (Annex Figure-1).

**Figure-1:** Event Study results by quarters

               **(A): Without New-Delhi**              **(B): With New-Delhi**

---

[13] Reference category is men in non-scheme states surveyed in Q4 of 2019.

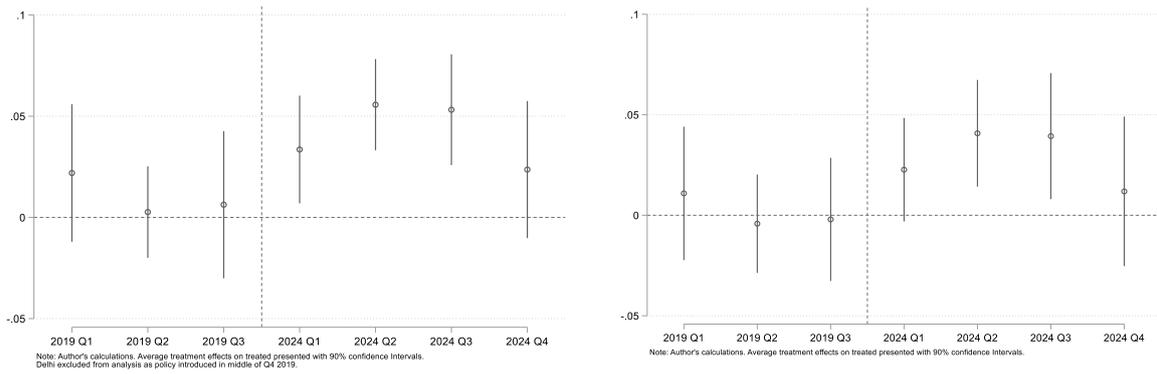

*c. Reallocation effects: Unpaid domestic labour and discretionary time for socialisation and leisure*

We also hypothesise that free bus scheme may increase discretionary time for women. Our triple difference estimates reveal that the scheme led to a higher likelihood of women reporting any socialisation (3 pp.) and engagement in at-least 1 hour of socialisation (2 pp.) in a day. Further, we also find that the probability of women reporting at-least 2 hours of leisure increased by about 4 pp., while there was no change in time spent on unpaid domestic labour. These findings suggest that free bus scheme allows women to reallocate time towards socialisation and leisure without incurring additional burden of unpaid domestic labour (Figure-2).[14]

**Figure-2:** Bus scheme effects on socialisation, leisure and chores & care work

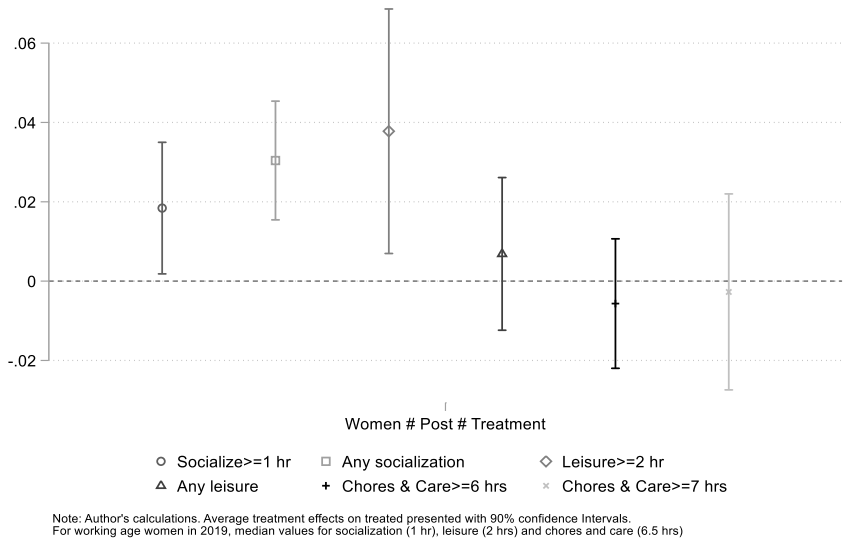

---

[14] We use median time spent by working-aged women in 2019 on respective activities as thresholds. These are 1 hour for socialisation, 2 hours for leisure and 6.5 hours for chores and care work. For chores and care work, we find no significant effects by level of education. Results available on request.

### d. Heterogeneous Treatment Effects

#### d.1. State level heterogeneity

Our findings show that the effects of the bus scheme are more pronounced in Punjab and Tamil Nadu, where the scheme facilitated improvements in WLFP both on the extensive and intensive margins. In Punjab, the probability of women's participation in paid work increased by about 9 pp., with the average daily duration of paid work increasing by 53 minutes. The corresponding estimates for Tamil Nadu are 5 pp. and 10 minutes, respectively. For Telangana, the effects were only observed on the intensive margin with an average increase of 47 minutes a day on paid work (Figure-3). These results should be interpreted in context of different baseline levels in 2019, where Punjab's WLFP rate was 17 percent in comparison to 38.5 percent in Tamil Nadu. Similarly, the baseline daily paid work duration for women was 58 minutes in Punjab and 153 minutes in Tamil Nadu.

The state wise heterogeneity effects reveal two crucial insights. One, the bus scheme is effective in improving WLFP rates across varied socio-economic contexts. Punjab and Tamil Nadu have historically featured at the bottom and top end of the state-level WLFP distribution, and both observe positive and significant effects of the scheme. This indicates that the bus scheme can potentially increase women's workforce participation in laggard and leading states. Second, the results are suggestive of the importance of exposure duration in realising the gains from such a scheme. Both Punjab and Tamil Nadu implemented the scheme in 2021 and had exposure to the scheme for close to three years before the 2024 survey. In contrast, states with recent rollouts like Karnataka and Telangana show no extensive margin effects. We posit that the effects of the scheme may not be instantaneous and may instead be mediated through gradual lowering of barriers related to women's mobility and workforce participation. Thus, broadening of WLFP may involve a gestation period. Next, we discuss the potential constraints placed on women's workforce participation by patriarchal norms and mobility restrictions.

**Figure-3:** Bus scheme effects by states

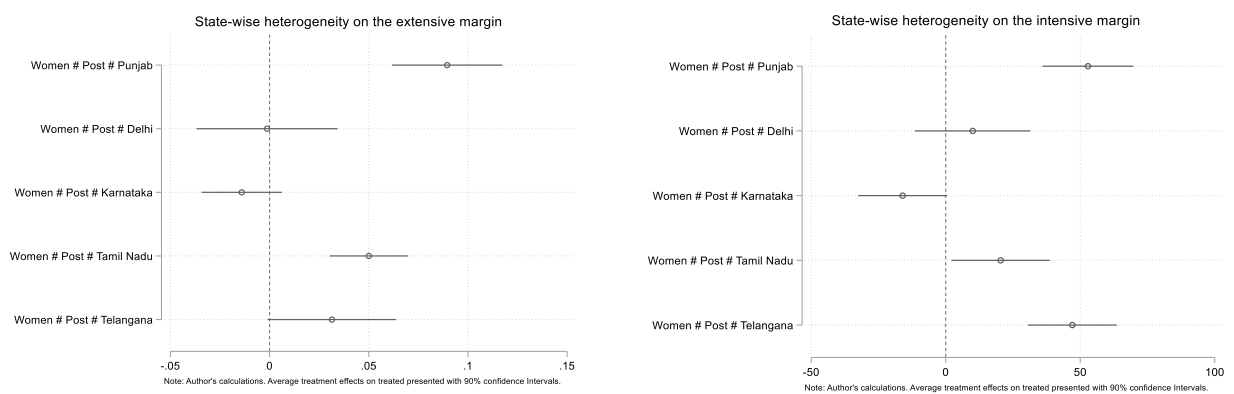

**Note:** The extensive margin results for Telangana and the intensive margin results for Karnataka are both insignificant (p value=0.11).

*d.2. Heterogeneous effects by patriarchal norms and mobility restrictions*

We now present the findings from our restricted triple difference specification for women (equation 2). In this specification, we alternatively interaction standardised measures of (a) patriarchy and (b) restrictions on women's ability to visit markets, both of which are computed at the district level from the DHS survey. The coefficients of interest are positive and significant both on the extensive and the intensive margins of WLFP. This suggests that the bus scheme generates disproportionately larger gains for women residing in districts characterised by higher patriarchy and mobility restrictions. On the extensive margin, one standard deviation increase in the district level patriarchy index and mobility restrictions generate treatment effects ranging from 2.4 pp. to 4.5 pp. On the intensive margin, these effects range from 9 to 25 minutes of additional daily paid work (Figure-4).

The findings indicate that the free bus scheme has a higher impact in more patriarchal settings. We posit that this provision does more than simply lowering the transit costs and potentially generates social benefits through increased ridership and shared occupancy, which lower the safety barriers and reservations against women's mobility. Thus, the policy has the potential to enhance women's bargaining power, which allows them to renegotiate time allocation for economic activities outside the home.

*Figure-4*: Bus scheme effects by patriarchal norms and mobility restrictions

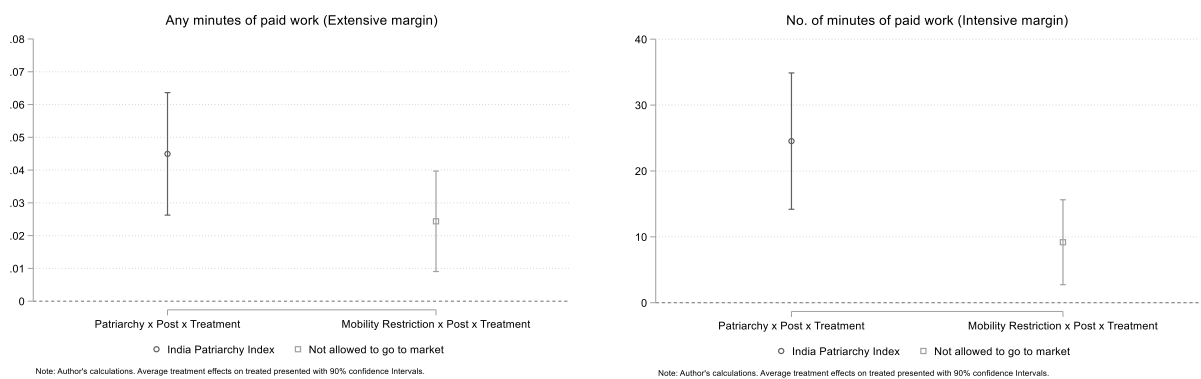

## 5. Conclusion and discussion

Owing to the disproportionately higher burden of unpaid domestic work, women's commute patterns are vastly different from those of men, with higher reliance on public transport (Perez, 2019). Evidence shows that women's mobility is also curtailed socio-cultural norms, limited agency in economic decision-making and safety concerns (Borker, 2022; Kondylis, et al., 2020; Heath, et al., 2024). These factors dictate that women's workforce participation is closely intertwined with access to affordable public transport (Alam, Cropper, Dappe, & Suri, 2021; Perez, 2019). In recent years, multiple state governments in India have introduced free bus schemes for women as a policy to facilitate higher workforce participation. This paper evaluates whether these schemes have improved WLFP and enhanced women's welfare.

Using a triple difference estimation strategy coupled with an event design framework, we find credible evidence of free public bus scheme for women increases the likelihood of their participation in paid work (extensive margin) by 3 pp. and average daily duration (intensive margin) by 16 minutes. As against the 2019 baseline for women, these estimates represent an increase of 13 percent on the former and 20 percent on the latter. Our findings also support this hypothesised mechanism through increase in women's likelihood of undertaking any work-related travel by 5 pp. The event study framework reveals no evidence of any prior trends with women's ATT for the three quarters preceding the scheme rollout being statistically indistinguishable from zero, while the those for quarters after the scheme's introduction being positive and significant.

These effects are not uniform across states and are mainly concentrated in Punjab and Tamil Nadu, two states that historically represent the laggard and leader states in Indian with respect to WLFP. We also note that in comparison to states like Karnataka and Telangana which implemented this policy in 2023, Punjab and Tamil Nadu rolled out these schemes relatively earlier in 2021. We argue that time may be needed to coordinate on job openings at proximate sites and create travel groups to sufficiently lower safety barriers before one sees an improvement in WLFP. Thus, broadening of women's workforce participation may involve a gestation period with the effects of such a policy becoming prominent with time. While we are unable to formally test this hypothesis, concentration of these effects among early adopters appears to be consistent with the notion that sustained exposure to the scheme may be needed to lower barriers related to norms, which are typically rigid, before seeing improvements in WLFP. We note that even though New-Delhi was the first to adopt a free bus scheme in 2019, we see no significant effects on the outcomes of interest. Although the available data does not allow us to explore why this may be the case, news reports suggest that New Delhi's unique Union Territory status which involves overlapping of administrative jurisdictions between the Union and the state government, and their frequent disagreements may have interfered with implementation of the scheme.[15]

Importantly, our findings reveal that the effects of the bus scheme are disproportionately higher for women residing in more patriarchal districts with stronger restrictions on women's mobility. As discussed previously, evidence shows social norms and safety concerns are a crucial determinant of women's mobility and influence the decision to participate in paid work. In areas where regressive social norms serve as a binding constraint against women's mobility and their workforce participation, the free bus scheme for all women encourages higher ridership which provides a social cover for even more women to travel, thereby starting a virtuous cycle. In contrast, for districts where such norms are less of a concern, the binding constraints on higher WLFP may be unrelated to free transport and be linked to other factors such as skill-mismatch or wage bargaining.

---

[15] https://www.business-standard.com/article/news-ians/centre-aap-in-tussle-over-free-rides-scheme-119060601442_1.html (last accessed on 5th April 2026).

Simultaneously, we also find that the time saved by women was allocated to more socialisation and leisure with no increase in the burden of unpaid household labour. This suggests that the "time-tax" on women may have reduced with them choosing public bus transport over other inefficient or unsafe mobility options. Together, these effects foster creation of social capital, which can have positive spillover effects that are instrumental in reducing the burden of socio-cultural norms and enhance women's bargaining power. The resulting social benefits can have far-reaching consequences that go well beyond improvements in WLFP. However, these are mere conjectures. Measuring safety perceptions and social capital and testing if they are associated with subsidised transport is an important area of future research.

Our findings have significant policy implications. One, our findings show that the state governments can leverage the free bus scheme for women as a tool for inclusive economic growth and development. Two, the scheme's effectiveness in improving WLFP in Punjab and Tamil Nadu, two states on the opposite spectrum of state level WLFP distribution, demonstrates that the policy is not an exclusive lever for states with low WLFP, but can also generate significant benefits for states where women's workforce participation in already well integrated with the overall labour markets. Three, concentration of the effects among early adopters highlights the importance of sustained implementation and requirement of a gestation period before labour market outcomes improve. This suggests that short-term evaluations may understate the potential long-term gains, which requires objective assessments of the scheme over a suitable time horizon. Four, stronger policy effects for more patriarchal and mobility restricted districts is suggestive of the importance of complementary investments in improving safety infrastructure, fleet expansion, and route rationalisation to maximise the marginal gains from the scheme. Without these complementary measures, the quality of the services may be compromised in ways that it makes the bus scheme redundant. Recent reports reveal that this is already a serious concern and needs to be addressed. For example, a report from the CAG, India's foremost auditing authority, shows that the size of the public bus fleet in New-Delhi has reduced in recent years, with lack of adequate maintenance leading to frequent breakdowns and causing overcrowding and rising safety concerns (Jamba, Devaraj, & Kanuri, 2025). Free transit schemes for women, which constitute roughly half of India's 1.4 billion population, can thus play a transformative role in enhancing women's economic and welfare outcomes.

# Annex

Table A1: State-wise trends in LFP

|  | 2019 | | | 2024 | | |
| --- | --- | --- | --- | --- | --- | --- |
| **States** | **MLFP** | **FLFP** | **Difference** | **MLFP** | **FLFP** | **Difference** |
| Jammu & Kashmir | 74.5 | 14.5 | 59.9*** | 83.9 | 10.7 | 73.2*** |
| Himachal Pradesh | 76.5 | 29.6 | 46.9*** | 82.1 | 25.1 | 57.0*** |
| Punjab | 84.4 | 17.3 | 67.0*** | 87.8 | 26.0 | 61.8*** |
| Uttarakhand | 72.5 | 11.4 | 61.1*** | 78.4 | 13.9 | 64.5*** |
| Haryana | 77.3 | 15.2 | 62.1*** | 85.5 | 18.4 | 67.1*** |
| Delhi | 82.3 | 15.6 | 66.7*** | 89.0 | 15.4 | 73.6*** |
| Rajasthan | 75.0 | 20.7 | 54.3*** | 80.6 | 31.4 | 49.2*** |
| Uttar Pradesh | 70.5 | 11.3 | 59.2*** | 81.8 | 16.2 | 65.6*** |
| Bihar | 82.3 | 7.4 | 74.9*** | 83.6 | 14.0 | 69.7*** |
| Sikkim | 79.8 | 36.6 | 43.2*** | 86.7 | 35.8 | 51.0*** |
| Arunachal Pradesh | 40.2 | 12.5 | 27.8*** | 90.2 | 14.3 | 75.9*** |
| Nagaland | 58.9 | 24.5 | 34.4*** | 53.5 | 14.9 | 38.6*** |
| Manipur | 75.4 | 44.8 | 30.7*** | 85.7 | 49.7 | 36.0*** |
| Mizoram | 88.6 | 31.9 | 56.8*** | 92.5 | 34.7 | 57.8*** |
| Tripura | 84.6 | 20.9 | 63.7*** | 89.3 | 20.7 | 68.6*** |
| Meghalaya | 59.7 | 26.3 | 33.4*** | 87.6 | 29.2 | 58.4*** |
| Assam | 79.5 | 13.3 | 66.2*** | 89.1 | 17.1 | 72.0*** |
| West Bengal | 85.3 | 19.5 | 65.8*** | 89.1 | 24.2 | 64.8*** |
| Jharkhand | 69.8 | 11.3 | 58.4*** | 86.7 | 24.4 | 62.3*** |
| Odisha | 79.1 | 17.7 | 61.4*** | 83.9 | 19.4 | 64.6*** |
| Chhattisgarh | 75.9 | 33.8 | 42.0*** | 81.8 | 39.2 | 42.7*** |
| Madhya Pradesh | 75.8 | 25.9 | 49.9*** | 88.5 | 26.7 | 61.7*** |
| Gujarat | 87.5 | 32.8 | 54.6*** | 89.8 | 34.6 | 55.2*** |
| Maharashtra | 84.1 | 35.5 | 48.5*** | 88.2 | 37.7 | 50.5*** |
| Andhra Pradesh | 85.9 | 36.8 | 49.1*** | 88.2 | 41.5 | 46.7*** |
| Karnataka | 80.6 | 27.3 | 53.3*** | 91.6 | 33.1 | 58.6*** |
| Kerala | 80.9 | 26.2 | 54.7*** | 81.8 | 30.3 | 51.6*** |
| Tamil Nadu | 87.4 | 38.5 | 48.8*** | 89.4 | 41.9 | 47.5*** |
| Telangana | 88.4 | 47.4 | 40.9*** | 88.1 | 46.9 | 41.2*** |
| **India** | **79.2** | **23.4** | **55.7*** | **86.2** | **27.3** | **58.9** |

Note: Authors' calculations

**Annex Figure-1:** Event Study, using months instead of quarters

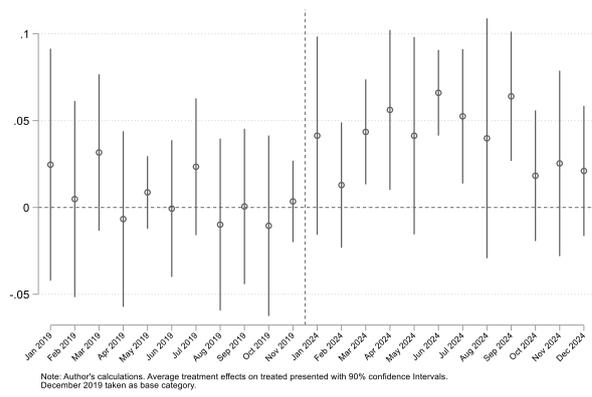

Note: Author's calculations. Average treatment effects on treated presented with 90% confidence Intervals.
December 2019 taken as base category.